%% file: main.tex
\documentclass[sigconf]{acmart}

\renewcommand\footnotetextcopyrightpermission[1]{} 
\AtBeginDocument{%
	}

\acmConference[Conference acronym 'XX]{Make sure to enter the correct	conference title from your rights confirmation email}{June 03--05, 2018}{Woodstock, NY}
\acmISBN{978-1-4503-XXXX-X/2018/06}

\usepackage{booktabs} 

\graphicspath{{figure/}{figures/}}

\usepackage{shortcuts}
\usepackage{lipsum}
\usepackage{listings}
\usepackage[switch]{lineno}
\usepackage{comment}
\usepackage{multirow}
\usepackage{makecell}
\usepackage{tablefootnote}
\usepackage{array}
\usepackage{tikz}
\usepackage{boldline}
\usepackage{url}
\usepackage{hyperref}
\usepackage{framed}
\usepackage{scalerel}
\usepackage{subcaption}
\usepackage{enumitem}
\usepackage[scaled=0.79]{beramono}
\usepackage[T1]{fontenc}
\usepackage{dirtree}
\usepackage{booktabs}
\usepackage{pifont}
\usepackage{wrapfig}
\usepackage{relsize}
\usepackage{fontawesome5}   
\usepackage{algorithm}
\usepackage{algpseudocode}

\setlist{leftmargin=2em}

\newcommand*\OK{\ding{51}}
\newcolumntype{H}{>{\setbox0=\hbox\bgroup}c<{\egroup}@{}}

\setcopyright{none}

\begin{document}
\title{A Longitudinal Study of Android Apps Signing Key Protection}
	


\author{Mark Huasong Meng}
\affiliation{%
	\institution{University College Dublin}
	\country{Ireland}
}

\author{Qing Zhang}
\affiliation{%
	\institution{JD Group}
	\country{China}
}

\author{Weirao Lu}
\affiliation{%
	\institution{JD Group}
	\country{China}
}

\author{Chunyang Chen}
\affiliation{%
	\institution{Technical University of Munich}
	\country{Germany}
}

\renewcommand{\shortauthors}{M. H. Meng et al.}

\begin{abstract}
	\input{sections/abstract2}
\end{abstract}
%
%

\maketitle
\pagestyle{plain}

\input{sections/intro}
\input{sections/background2}

\input{sections/methodology2}

\input{sections/findings}
\input{sections/exploit}
\input{sections/broader_impact}

\input{sections/disclosure}

\input{sections/discussion2}
\input{sections/related_work}
\input{sections/conclusion}
\input{sections/data_availability}
\bibliographystyle{ACM-Reference-Format}
\bibliography{references} 

\end{document}

%% file: sections/abstract2.tex

Android app signing relies on developer-managed credentials, making secure key protection essential for the integrity of the software supply chain.
A recent platform key leakage incident involving two major OEM manufacturers demonstrates that even robustly designed signing mechanisms can be compromised due to developers' oversight. 
In this work, we conduct a longitudinal ecosystem study to characterize this threat by mining public repositories for Android signing credentials, recovering compromised keys via exposed passwords, and matching them against signatures from over 4,000 apps collected from major stores and OEM system images.
Our analysis identifies 5,673 compromised keystores on GitHub and 26 unique certificates linked to 278 real-world apps.
These include 26 third-party apps in public app stores and 252 preinstalled apps from seven manufacturers, collectively affecting over 10 billion users. We demonstrate the practical exploitability of these leaks through a proof-of-concept app replacement attack and identify spillover risks in non-smartphone platforms, including a popular automotive head-unit platform installed in over 1,100 vehicle models. Our results reveal that signing-key mismanagement is a systemic risk, underscoring the need for a more rigorous key-management support in Android release engineering and distribution infrastructures.

%% file: sections/intro.tex
\section{Introduction} \label{sec:intro}

App signing serves as the primary security mechanism to protect the integrity and authenticity of applications (apps) after they are released by developers.
As in most operating systems (OSes), Android is designed with an app signing framework that forms a critical component of its security architecture~\cite{google2023sign}. 
When developers release their apps, Android requires all apps to be digitally signed with the developer's cryptographic keys before distribution.
During installation, the Android OS verifies the app’s signature, ensuring that only apps signed by the same developer can replace or upgrade existing versions. 
This mechanism not only safeguards app integrity but also enables advanced features such as permission-based privilege sharing between apps released by the same developer. 

In recent years, critical vulnerabilities like Log4Shell have pushed software supply chain security to the forefront of cybersecurity concerns~\cite{roberts2024history,ibm2024log4shell,nist2021cve}. 
As app signing constitutes a critical component within the software supply chain, its reliability directly impacts the security of the entire Android ecosystem.
Unfortunately, the Android app signing mechanism has indeed exposed several significant vulnerabilities in the past~\cite{wang2019characterizing}.
Notable vulnerabilities, such as the Master Key vulnerability (CVE-2013-4787)~\cite{nist2013cve} and the Janus vulnerability (CVE-2017-13156)~\cite{nist2017cve}, have exposed fundamental design flaws in the signing mechanism. 
These issues were not comprehensively addressed until the release of Android 9, after which the app signing mechanism has not exhibited severe flaws, with remaining risks largely attributed to whether developers follow secure development practices and adequately protect their signing keys.

In this context, the security implications of signing-key compromise are not limited to apps distributed through Google Play. In practice, many Android apps are obtained outside Google Play's security controls, including apps released in alternative regional markets, apps on devices without Google Play services, and OEM-preinstalled apps that are not listed in the Play Store. In these settings, Google Play's store-side protections~\cite{google2026protect} do not uniformly apply, making developer- and OEM-side signing-key protection particularly important and leaving greater room for the consequences of signing-key mismanagement to surface in real-world deployments.

In this work, we investigate the security landscape of Android app signing with a particular focus on the risk of signature leakage. 
Our research is motivated by a recent OEM platform key leakage incident~\cite{newman2022android}, which prompted us to examine whether developers' mismanagement of signing keys could pose a broader threat beyond a single app or even to the entire Android ecosystem, including smart home platforms and on-vehicle head units.
Moreover, our concerns are elevated by findings in recent literature~\cite{wang2019characterizing} indicating that some amateur Android developers use publicly known keys to sign their applications. Once such keys are leaked, attackers can craft malicious apps bearing authentic signatures.
Given Android's unique openness and its substantial market share, there exists a strong demand for a comprehensive assessment to strengthen confidence in the security of the Android software supply chain.

To address these concerns, we conduct a large-scale empirical study of the Android app signing mechanism. 
Specifically, we longitudinally crawl public repositories to collect signing credentials and identify compromised keys by searching for associated passwords within the same repositories. 
In parallel, we gather over 4,000 apps from major app stores and OEM system images.
By matching app signatures with the compromised signing keys, we evaluate the potential impact of signing key leakage issues and reveal the security landscape of the app signing mechanism in the Android ecosystem.
Finally, we design two manipulation strategies and implement a proof-of-concept attack on real devices to demonstrate the exploitability of our findings. 
We also discuss the broader risks posed by signing key leakage, extending beyond smartphone devices to other Android-based platforms.
Among many interesting results and observations, the following are the most prominent:

\begin{itemize}
\item \textbf{Mismanagement of app signing keys is prevalent in the Android ecosystem}. \noUniqueKsCracked unique keys are found compromised on public repositories on GitHub. 
\item \textbf{\noofVulnerableAppsThirdParty apps released on mainstream app stores are found to be vulnerable} due to the compromised signing keys. These apps have aggregated over \noofInstallsThirdParty installs across seven app stores.
\item \textbf{\noofVulnerableAppsPreinstalled preinstalled apps are found signed with 13 compromised keys on devices from 7 manufacturers}, posing severe threats to the users' devices if the malicious app misuses privileged permissions or deploys a Denial-of-Service (DoS) attack.
\item \textbf{A proof-of-concept attack} that can be installed to replace a benign real-world app on our tested device, demonstrating the exploitability of the signing key leakage issues.
\item \textbf{A signing key for a third-party automotive head unit platform found compromised}, which has been installed on over 1,100 vehicle models since 2019. This indicates that the broader risks have extended beyond smartphone devices to other Android-based platforms.
\end{itemize}

To the best of our knowledge, this is the first systematic study of potential leakage of app signing keys in the Android ecosystem at a large scale and longitudinally. Our findings should alert both developers and OEM manufacturers to comply with rigorous security practices throughout the entire development life cycle and urge the industry to adopt a more secure key management solution. 

%% file: sections/background2.tex

\section{Background}
\label{sec:background}

\begin{figure}[t!]
	\centering
	\vspace{-0pt}
	\includegraphics[trim=0cm 0cm 0cm 0cm, clip=true,width=1\linewidth]{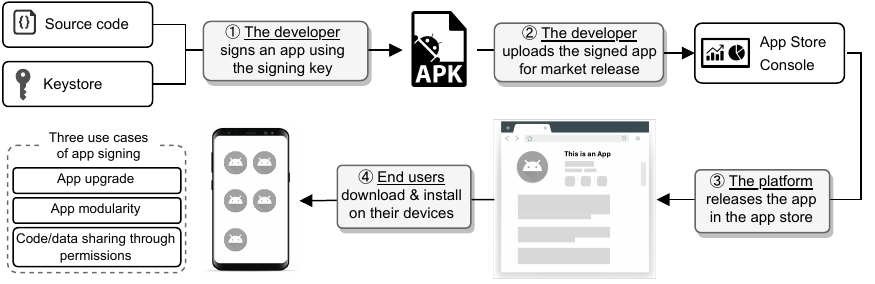}
	\vspace{-0.2cm}
	\caption{The role of app signing in the life cycle of an Android app}
	\label{fig:signing}
	\vspace{-0.1cm}
\end{figure}

\subsection{Android App Signing \& Trust Relationships}

Publishing an Android app is not only about transforming functionality requirements into source code; developers are also expected to protect their apps' integrity and genuineness from spoofing attacks.
In the Android ecosystem, app signatures are used to prove the data integrity of the app package and ownership of its developer~\cite{google2023sign}.

The process of app signing is depicted in Figure~\ref{fig:signing}.
Specifically, a developer needs to generate a key pair consisting of a private key and a public key, create a certificate that specifies developer identity information, and sign the app package with the private key and certificate. 
The key pair is usually saved in Java keystore files (with extensions \texttt{.jks} or \texttt{.keystore}).
When a developer finishes the implementation of an app, 
the signing material is then embedded into the installation package file (\texttt{.apk} file, we refer to it as APK in the remainder of this paper) during the release process (step 1)~\cite{google2023application}.
In practice, developers can resort to development tools (e.g., Android Studio) to generate keys and sign an app through a graphical user interface (GUI) with a few clicks.
Corporate developers who simultaneously develop and maintain multiple apps can also configure the signing process into the development pipeline, for example, by specifying the signing key and password in a ``\texttt{Build.gradle}'' file in the project. 

After the signing, the developer uploads the signed package to the app store console provided by the release platform, e.g., Google Play App Store (step 2). The release platform will publish the app after performing a few integrity checks (step 3) and then make it public for end users to download and install on their devices (Step 4).

The OS on the users' devices checks the signature during the installation, and only allows app upgrades if both the new and old versions of an app are signed by the same developer.
Signature continuity can also mediate certain trust relationships between apps signed by the same developer, including signature-level permission sharing~\cite{google2025permissions}. Consequently, leakage of a signing key can enable an attacker to distribute forged apps that inherit the developer's signing identity.

\subsection{Threat Model and Scope}

In this work, we study \emph{developer-side exposure} of Android app signing credentials in public code repositories. Our focus is not on cryptographic flaws in Android signature schemes, but on operational key-management failures that expose keystore files and associated passwords during development and release engineering. This threat is significant because Android relies on signature continuity to authenticate app updates and to support same-signer trust relationships. In practice, signing keys are long-lived and difficult to replace without disrupting an app's normal lifecycle.
Google recommends that private signing keys should remain valid for at least 25 years~\cite{google2023sign}.
App developers are also responsible for securing their  keys and the containing keystore files.
Google also suggests that developers maintain the keystore files in hardware or services that require multiple factors of authentication to maximize signing-key protection~\cite{google2024practices}.

We consider an attacker who can access artifacts unintentionally exposed in a public repository, including keystore files and related build configuration files. The attacker is assumed to inspect the same repository for information that enables keystore recovery, such as plaintext passwords or file references to password material. Under this threat model, if a signing key can be recovered, the attacker may produce a forged app package  (i.e., an \texttt{.apk} file) bearing the victim developer's signing identity. This forged package could then be used to attempt a malicious update, impersonate a same-signer app, or abuse signature-based trust relationships. The exact impact depends on where the affected app is deployed and what privileges it holds. 

Our analysis covers two major app deployment contexts. The first is third-party apps distributed through mainstream app stores. The second is preinstalled or system apps on OEM devices\footnote{We refer to ``system apps'' as apps that are installed in system directory (e.g., ``\texttt{/system/app/}'', ``\texttt{/system/priv-app/}'', etc) and therefore might be given certain privileged permissions.}, where compromise may be especially serious because signature-based trust can interact with privileged permissions. We also discuss broader implications for other Android-based platforms when relevant keys are reused outside the smartphone ecosystem.

It is worth noting that our analysis scope excludes cases where a keystore is present online but cannot be linked to a real-world app, as well as cases where the signing identity is observed in an app but the corresponding keystore cannot be recovered (i.e., password not found). For that reason, our study separates \emph{potential leakage} from \emph{confirmed compromise}.
Specifically, matching a repository-exposed signing identity (i.e., certificate fingerprint) to a released app indicates potential exposure, whereas successful recovery and use of the signing key demonstrates practical exploitability.

%% file: sections/methodology2.tex

\begin{figure*}[t]
	\centering
	\vspace{-0pt}
	\includegraphics[trim=0cm 0cm 0cm 0cm, clip=true,width=0.6\linewidth]{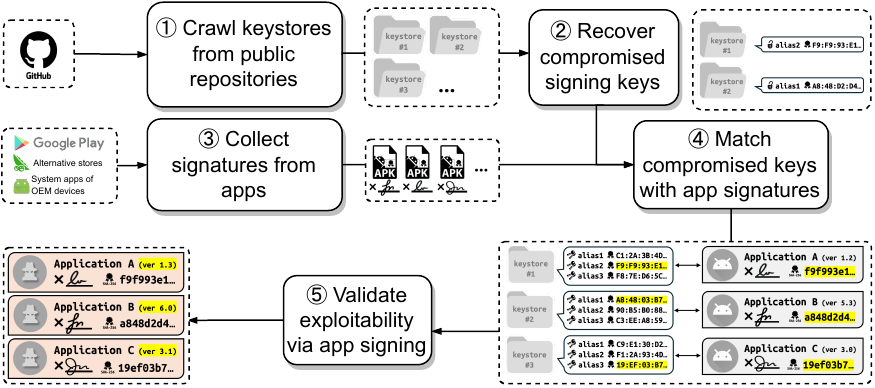}
	\caption{Overview of our methodology, including repository-side compromise identification, ecosystem-side signature matching, and exploit  deployment}
	\label{fig:overview}
	\vspace{-0.1cm}
\end{figure*}

\section{Study Methodology}

We present the details of our study in this section.
We first introduce the study design and define four research questions that we focus on in this paper. We then outline our methodology as a three-stage pipeline and briefly introduce different steps. Finally, we present the detailed methodology.

\subsection{Study Design} \label{sec:study_design}

In this paper, we focus on the following research questions:
\begin{itemize}
	\item \textbf{RQ1 Signing Key Mismanagement.} 
    How many Android app signing keys are hosted on public repositories on GitHub? Can we find any keys compromised due to password leakage?
	\item \textbf{RQ2 App Signature Leakage.} 
    Have the compromised keys, if any, been used for signing real-world apps on Android devices? How many apps are vulnerable due to the mismanagement of app signing keys?
	\item \textbf{RQ3 Potential Impact.} 
    What is the potential impact of the signing key leakage accident? How many users or devices are affected?
	\item \textbf{RQ4 Exploitation.} 
    For the signing key leakage accidents, does there exist a practical exploit methodology that threatens users and devices?  
\end{itemize}

\subsection{Approach Overview} \label{sec:overview}

Our study follows a three-stage pipeline, as illustrated in Figure~\ref{fig:overview}.
In \textbf{Stage 1} (\emph{repository-side compromise identification}), we first crawl keystore files for Android app signing from public GitHub repositories (step 1) and then attempt to recover their signing keys by inspecting the same repositories for exposed passwords and verifying keystore access (step 2). We present more details of this stage in Section~\ref{sec:crawl-and-decrypt-keys}.

In \textbf{Stage 2} (\emph{real-world app matching}), we collect signatures from released Android apps, including both third-party apps from mainstream app stores and preinstalled system apps from OEM devices (step 3), and match them against the certificate fingerprints of the recovered signing keys (step 4). A match indicates that a compromised signing key has been used by a real-world app. We detail stage 2 in Section~\ref{sec:signature-matching}.
The security implications of these matched cases are analyzed later in Section~\ref{sec:risk}.

Finally, in \textbf{Stage 3} (\emph{exploitability validation}), we use the recovered keys to sign forged apps and conduct a spoofing attack. Specifically, we assess whether the forged apps can be accepted as legitimate updates (step 5). 
A successful spoofing attack demonstrates the practical exploitability of a previously recovered signing key. All apps signed by that key are exposed to malicious manipulation. We detail this stage in Section~\ref{sec:exploit}.

With the above three stages, our pipeline connects repository-side evidence to real-world app exposure and practical exploit validation.

\subsection{Repository-Side Compromise Identification} \label{sec:crawl-and-decrypt-keys}

\subsubsection*{\textbf{Crawling Keystores from Public Repositories}}
Leakage of an app-signing key may arise from multiple sources, including unauthorized theft or inadvertent public exposure of the corresponding keystore file. In this work, we focus on the latter. Collecting keystores obtained through other attack channels at scale would be both technically difficult and ethically inappropriate, whereas publicly exposed repository artifacts provide a measurable and reproducible source of developer-side signing-key mishandling.

We therefore target public GitHub repositories, which constitute a large platform for hosting Android source code. Because exhaustive extension-based retrieval is constrained by GitHub search limitations, we design a three-phase collection approach using RESTful API requests:
 
\begin{enumerate}[wide]
	\item \textbf{Identifying the repositories}. 
	Since our goal is to collect keystores used for Android app signing, we first identify repositories that are likely to host Android projects. We require candidate repositories to contain a build configuration file with the ``\texttt{.gradle}'' extension~\cite{google2025configure}. To further improve precision, we search for signing-related keywords derived from the \texttt{signingConfigs} section of a typical ``\texttt{Build.gradle}'' file, including  ``keyAlias'', ``storeFile'', ``storePassword'', and ``keyPassword''~\cite{google2023sign}. To avoid missing repositories due to GitHub result limits, we partition the search space by constraining the size of the target \texttt{.gradle} files and iterating over file size ranges from zero KB to 100 MB.
	
	\item \textbf{Collecting keystores from repositories}. 
	For each identified repository, we download the repository archive as a ZIP file\footnote{The zip file can be directly downloaded through the formatted URL \url{https://github.com/<username>/<repository-name>/archive/<branch-name>.zip}.} and inspect its file list for keystore artifacts with the ``\texttt{.jks}'' or ``\texttt{.keystore}'' extension. Whenever such a file is found, we add it to our candidate keystore set for further validation.
	
	\item \textbf{Validating the keystores}. 
	Not all files with these extensions are valid keystores. Some are placeholders, such as files containing strings like ``redacted'' or ``replace me with your keystore''. We therefore validate each collected file using Java \texttt{keytool} and retain only those that contain at least one private-key/certificate pair suitable for Android app signing.
\end{enumerate}

Because our goal is to characterize the evolving landscape of Android signing-key exposure, we perform this collection longitudinally rather than as a one-time crawl. Our keystore collection begins in \startDate and is repeated monthly until \collectTill.


\subsubsection*{\textbf{Recovering Compromised Signing Keys}}
Releasing the keystore files on a public repository does not directly imply a security breach unless there is a way to efficiently access the private keys embedded in these keystores.
While our work does not aim to crack these keystores, we instead resort to a simple but straightforward approach to search for the password in plaintext within the same repository. 
This is motivated by occasional reports in the past that some developers hardcoded the passwords to access their keystore files and private keys in the configuration file, i.e., ``\texttt{Build.gradle}''~\cite{hussain2019storing}. 
We regard the key signing process in the Android ecosystem as centrally managed in the configuration files, and accordingly search for them in the same repository that stores the keystore files. 
Specifically, we revisit these configuration files, parse the passwords in their signing configuration section (i.e., the \texttt{storePassword} and \texttt{keyPassword} elements in \texttt{signingConfigs}), and treat the keystores as \emph{recoverable} if any of two conditions are satisfied in the value of the passwords: (1) it is hardcoded plaintext, or (2) it is a file path and the file is publicly accessible in the same repository.

Next, we decrypt the keystore files using the passwords crawled from the same repositories. The correctness of a keystore password can be verified using \texttt{keytool}. In this work, we call the ``\texttt{keystore -list}'' command to list the signing keys in the keystore. The list of key aliases and certificates will be returned if the provided keystore password is correct.

\section{Identifying App Signature Leakage} \label{sec:signature-matching}

\subsubsection*{\textbf{Collecting Signatures from Apps}}
In this step, we aim to collect as many apps' APK files as possible to maximize our collection of apps' signatures.
To this end, our app collection starts with downloading the top 1,000 non-game free apps as of \startDate from the Google Play App Store, given that it is the official source of third-party apps in the Android ecosystem.
Due to some restrictions imposed by Google, we cannot directly download APK files to non-Android devices from its app store. Instead, we take the app ranking from the Google Play App Store, retrieve the package names of these apps, and download them through a third-party app repository named ``APKPure''.\footnote{Available at \url{https://apkpure.com/app}}
We also consider that the absence of Google Play services in China might limit the scope and impact assessment of our investigation, and therefore adopt a third-party app market with a large user base in China named ``Wandoujia''\footnote{Available at \url{https://www.wandoujia.com/}} as an alternative source of APK files to complement our app collection.
In the same way, we collect the identifiers of another 1,000 non-game free apps\footnote{Our collection does not contain duplicates in the two app stores. Thus, when we collect apps from Wandoujia, we skip the app if it has already been downloaded in the previous collection on the Google Play Store.} from the top list of the alternative app stores and download the APK files from APKPure to support lightweight retrieval of current APK versions. 

In addition to third-party apps circulating in the mainstream app stores, we also collect the preinstalled apps on different OEM devices.
These apps include preinstalled third-party apps that may be granted privileged permissions by the OEM manufacturers (also referred to as OEM vendors in the relevant study) and the system apps signed by the OEM manufacturers themselves, because leakage of a manufacturer key could pose broader ecosystem risk.
Specifically, we dump the preinstalled apps and the system apps\footnote{APKs located in \texttt{/system/framework/} directory, e.g., \texttt{framework-res.apk}.} from 11 devices, which are in different hardware models produced by Google Pixel and third-party OEM manufacturers including Honor, Huawei, Infinix, Meizu, OnePlus, Oppo, Samsung, Tecno, Vivo, and Xiaomi. Together, these manufacturers cover 87.5\% of the market share in the Android smartphone and tablet ecosystem at the time of the study being performed~\cite{appbrain2024top}.

For each APK file downloaded, we resort to a built-in tool of Android SDK named ``apksigner'' to retrieve the app signature.
Specifically, the certificate digest, signer's details, and public key digest of each app are collected for the next step of investigation.


\subsubsection*{\textbf{Matching Compromised Keys with App Signatures}} 
After collecting keystores and app signatures from the Internet, we then explore whether any certificates match between the two sets.
Specifically, we extract all certificate digests from the APK files, and compare each of them with the \emph{certificate fingerprints} recovered from the collected keystore binaries. We depict this process in step 4 of Figure~\ref{fig:overview}. 
If there exists a certificate from an app's signature that has the same digest value as the certificate fingerprint of the recovered keystores, we determine that the certificate is the one used to sign the app, and the corresponding keystore file is owned by the app developer.
In the same way, we regard this as a case of signing key leakage for the victim app and its developer(s). 

Unlike signing keys leaked on the Internet, certificates extracted from APK files usually contain lots of metadata about the developers, including countries, company and department names, and email addresses. Attackers can use this information to infer the market share and impact of a compromised signing key once matched with any app's signatures. 

%% file: sections/findings.tex
\section{Results and Analysis}

\subsection{RQ1: Signing Key Mismanagement}

\begin{figure}%
	\vspace{-0.2cm}
	\begin{minipage}[b]{0.49\linewidth}
		\centering
		\includegraphics[width=1\linewidth]{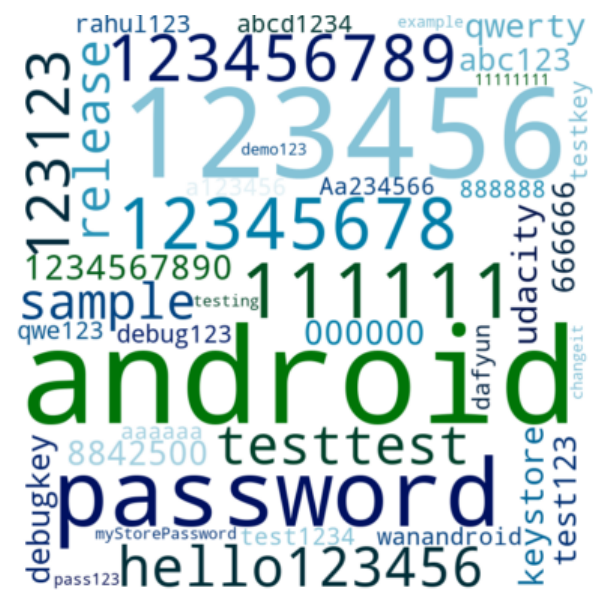}\\
		\subcaption{Keystore passwords}
	\end{minipage}%
	\hfill
	\begin{minipage}[b]{0.49\linewidth}
		\centering
		\includegraphics[width=1\linewidth]{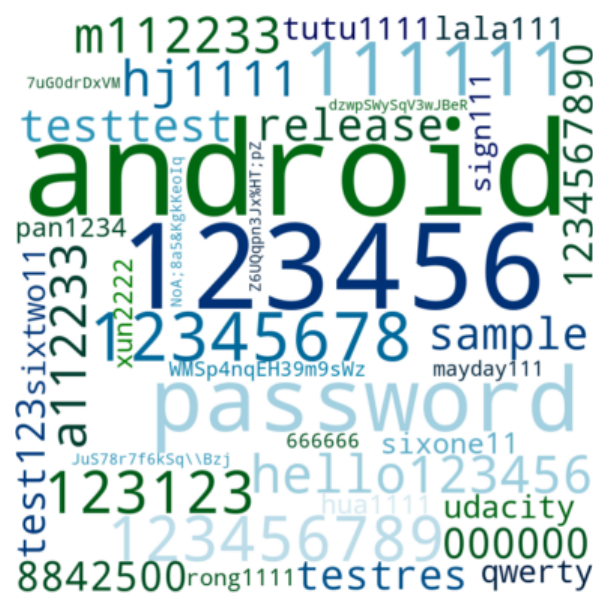}\\
		\subcaption{Signing key passwords}
	\end{minipage}%
	\vspace{-0.3cm}
	\caption{Word cloud pictures showing the distributions of passwords leaked on public repositories (only the 40 most frequently used passwords are shown due to space limit)}
	\label{fig:password}
\end{figure}

Through our large-scale longitudinal collection, we managed to collect \noAllKs keystore files from public repositories on GitHub. Among these \noAllKs keystores, we find \noUniqueKs unique keys (i.e., private key entries) by their certificate fingerprints (in the format of SHA-256 digests).

Among the \noAllKs keystores retrieved from online repositories, \noAllKsCrackedwPercent keystores are uploaded together with passwords in plaintext. We further remove duplicates by counting the unique fingerprint of the keystore files and recognizing \noUniqueKsCracked unique compromised keystores, in which \noUniqueKeys signing keys are identified. 

We also observe common patterns in the leaked passwords. Specifically, we find that ``\texttt{android}'', ``\texttt{password}'', and ``\texttt{123456}'' are the top three widely used passwords in both compromised keystores and signing keys. We illustrate the distribution of the leaked passwords in Figure~\ref{fig:password}.
Our statistics also show that the vast majority of compromised signing keys (\noUniqueKeysSamePwwPercent) share the exact same password with the embedding keystores.

\subsection{RQ2: Signature Leakage of Real-World Apps}

Our app collection managed to identify \noUniqueApps unique apps involving \noAppsSig distinct developer signatures.
\noUniqueAppswMultiSigs are found signed by multiple certificates, ranging from 2 to 11.
Among the \noUniqueAppswMultiSigs multiply signed apps, 252 (97.3\%) apps considered core system apps developed by Google and hardware vendor\footnote{We recognize the core system apps by package name patterns, e.g., ``\texttt{com.android.*}'' (153 apps), ``\texttt{com.qualcomm.*}'' (39), ``\texttt{com.google.*}'' (34), ``\texttt{android.*}'' (8), and ``\texttt{vendor.*}'' (3).}, which are signed by different OEMs during production. 

Our signature matching managed to find 26 compromised certificates involving \noofVulnerableApps apps and \noofVulnerableRepos repositories. 
Two compromised certificates are found in more than 10 apps. 
Specifically, a certificate with its fingerprint snippets ``\texttt{a40da80a59}'' is used by 27 apps. 
It is widely adopted in signing the compatibility test suit app\footnote{Identified by their package names, ``\texttt{com.android.cts.ctsshim}'' and ``\texttt{com. android.cts.priv.ctsshim}'', with signer email ``android@android.com''} in nine OEM devices.
The other one with its snippet ``\texttt{a6ef817bfd}'' exists in 17 apps. Through our analysis, we believe it is owned by one of the major players in the global mobile app market named Baidu because all the involved apps' package names start with ``\texttt{com.baidu.*}''.
We provide further analysis in Section~\ref{sec:risk}.

In addition to the compromised certificates, we also find a matching certificate without its password leaked on the Internet, which is found in a system app named ``\texttt{com.android.sdksandbox}'' in a device produced by Vivo. We believe it is owned by the OEM manufacturer because the signer email ``bbktel@bbktel.com'' identifies Vivo's parent company. Although the certificate is not classified as compromised due to the absence of a password, it is still at great risk of being cracked in a brute-force manner, and therefore, should be replaced with a new one in future releases of the affected apps.  

\subsection{RQ3: Impact and Risk Analysis} \label{sec:risk}

\input{tables/table_of_third_party_apps}

Based on the signature matching outcomes, we further explore the security impact of our findings. 

\subsubsection{Impact of Involved Third-Party Apps}
We first investigate the involved third-party apps that are publicly available on app stores. 
As shown in Table~\ref{tab:table-compromised-apps}, there are nine compromised keys that have been used for signing third-party apps circulating in public app stores. Each compromised key is owned by a different developer. 26 apps with their functionalities covering 10 categories are involved. 
We then visit the product pages of these 26 apps on app stores, and then count the number of downloads or installs based on the metadata available. 

In order to understand the real user volume, we resort to the statistics from as many app stores as possible. 
To this end, in addition to the Google Play Store and Wandoujia, we also searched the package name of the 26 apps on two additional public app stores, named Tencent app store and Baidu app store. 
As we can access the app stores that are privately owned by OEM vendors through our tested devices, for those OEM-owned app stores that display downloads or install numbers of an app, we additionally search the package names and record the number of downloads or installs.
As a result, we present a rough estimation of user scope by composing statistics of eight app stores, and sort the signing keys by the scale of impact in Table~\ref{tab:table-compromised-apps}.
As the table shows, there are at least 8 of the 26 apps that have gained over 1 billion installs. Two apps, namely ``BaiduMap'' and ``Ctrip'', have been downloaded over 10 billion times. Although these two apps mainly target Chinese users, such a large user base can still bring serious consequences once the leaked keys have been misused by attackers. 

\input{tables/table_of_preinstalled_apps_2}

\subsubsection{Compromised Certificates in Pre-Installed Apps}
Apart from apps from stores, the compromised keys are also widely used for signing pre-installed apps. 
Pre-installed apps signed with compromised keys may bring even bigger impact if they are system apps or even privileged apps\footnote{We refer to system and privileged pre-installed apps as the apps installed in a system directory (e.g., ``/system/app/''
and ``/system/priv-app/'') and granted ``preinstalled" privilege.} because these types of apps are allowed for certain privileged actions without users' awareness\footnote{Unlike dangerous permissions that request user consent, privileged permissions are typically granted automatically at installation time and come with system-level trust.}~\cite{google2025permission}. 

In this work, we find 13 compromised keys from the Internet that are used for signing pre-installed apps. 
Among these 13 certificates, we find that one certificate is used to sign third-party apps as pre-installed ones. 
Specifically, four pre-installed third-party apps are signed with that key on our tested devices from five OEM vendors. As shown in Table~\ref{tab:table-compromised-pre-apps}, all these four apps are developed by Baidu and are therefore signed using the same compromised key as what we found from the app stores (refer to Table~\ref{tab:table-compromised-apps}).
Four out of five OEM vendors choose to install these apps in privileged directories, i.e., ``\texttt{/system}'', ``\texttt{/vendor}'', and ``\texttt{/product}'', and consequently these apps become system apps or even privileged apps~\cite{google2025allowlisting,google2025what,sreelakshmi2025android}. For example, through our reverse engineering, we found that all four apps request access to device identifiers and the list of installed apps in their permission declaration. These data are considered highly sensitive user privacy and therefore, access to them has been banned from third-party apps since Android OS version 10~\cite{google2023privacy10} and 11~\cite{google2023privacy11}. However, these strict rules do not apply to system and privileged apps. These apps, even once replaced in a spoof attack, are still granted to freely access the sensitive data, resulting in a substantial privacy risk. 
Moreover, the exposure of privileged app certificate also brings a risk of permanent denial-of-service of the victim device, which we refer to as ``creating permission conflict'' and detail in Section~\ref{sec:exploit}.

\input{tables/table_of_aosp_key_usage_2}

\subsubsection{Misuse of AOSP Certificate}
In addition to the one third-party-owned certificate caught in pre-install apps, the remaining 12 compromised certificates are recognized as ``AOSP keys'' as they are released by Google in its Android Open Source Project (AOSP) as part of the ``testing kit''. 
These certificates are uploaded together with the private keys (as the pair of \texttt{.x509.pem} and \texttt{.pk8} files) in the \texttt{/product/security} directory. Google has underscored that the developers and vendors should \emph{never use them to sign any packages in publicly released images}~\cite{google2025signbuilds}. In other words, these 13 certificates represent ``exposure by design'' rather accidental leakage on the Internet. Unfortunately, we still find six vendors, including Google Pixel, that use the AOSP keys to sign their app packages (i.e., system components and vendors' in-house apps) in their OS images. We list these certificates and involved packages in Table~\ref{tab:table-aosp-key-usage}. 

The potential risk of misusing ``public'' certificates in signing the critical system components is far greater than that of signing third-party apps. Given that these system components or vendors' in-house apps are usually highly privileged with many sensitive permissions, once the attacker knows the exact certificate used, they can easily find the private key from the public AOSP projects. An attacker who wants to misuse the privileges may not even necessarily conduct a spoofing attack since an arbitrary app signed with the platform certificate can inherit the same set of permissions. 

%% file: tables/table_of_third_party_apps.tex
\begin{table*}[t]
\centering
\def\arraystretch{0.97}
\setlength{\tabcolsep}{2pt}
\caption{\label{tab:table-compromised-apps} List of third-party apps involved in signing key leakage}
\footnotesize
\resizebox{1\linewidth}{!}{%
\begin{tabular}{Hllllccccccccc}
\hline

&  &  &  &  & \multicolumn{8}{l}{\textbf{\makecell[lt]{Downloads \& Installs\textsuperscript{$\dagger$}}}} \\
\cline{6-14}
& \multirow{2}{*}{\textbf{\makecell[lb]{Certificate\vspace{-0.2em}\\Fingerprint}}} & & & & \multirow{2}{*}{\textbf{\makecell[lb]{Google\vspace{-0.2em}\\Play Store}}} & \multicolumn{3}{l}{\textbf{\makecell[lt]{Alternative Stores\textsuperscript{$\ddagger$}}}} & & \multicolumn{4}{l}{\textbf{\makecell[lt]{OEM Vendor Stores\textsuperscript{$\ddagger$}}}} \\
\cline{7-9}\cline{11-14}
\textbf{\#} &  & \textbf{Owner} & \textbf{Apps} & \textbf{Category} 
&  & \textbf{(A1)} & \textbf{(A2)} & \textbf{(A3)} & 
& \textbf{(O1)} & \textbf{(O2)} & \textbf{(O3)} & \textbf{(O4)}  \\
\hline
1  & \texttt{a6ef817bfd} & Baidu 
& \texttt{com.baidu.searchbox} 	& Tools 	& -- & 3.3B & 410M & 34.9M && 26.7B & 5.4B & 700M & 12.8B \\ \cline{4-14}
& & 
& \texttt{com.baidu.BaiduMap}	& Map	& 5M+ & 2.1B & 160M & 16.8M && 9.3B & 3.7B & 6.1B & 479M \\ \cline{4-14}
& & 
& \texttt{com.baidu.appsearch}	& Utility	& -- & 1.7B & 1.2B & 12.5M && -- & -- & -- & -- \\ \cline{4-14}
& & 
& \texttt{com.baidu.input}	& Input Method	& 5M+ & 270M & 100M & 9.2M && 700M & 230M & 270M & 100M \\ \cline{4-14}
& & 
& \texttt{com.baidu.browser}	& Browser	& -- & 170M & 74.6M & -- && 100M & 77.6M & 47.6M & 46.1M \\ \cline{4-14}
& & 
& \texttt{com.baidu.searchbox.lite}	& Tools	& -- & 160M & 99.2M & 8.5M && 2.1B & 940M & 1.5B & 1.1B \\ \cline{4-14}
& & 
& \texttt{com.baidu.searchbox.tomas}	& Tools	& -- & 5.1M & 300K & -- && 200M & 140M & 83.7M & 210M \\ \cline{4-14}
& & 
& \texttt{com.baidu.searchbox.senior}	& Tools	& -- & 113K & -- & 43K && -- & 540K & 674K & 640K \\
\hline
2 & \texttt{0a350c6fbf} & Kwai (Joyo Tech)
& \texttt{com.smile.gifmaker}	& Video	& -- & 2.4B & 1.6B & 62.9M && 29.4M & -- & -- & -- \\ \cline{4-14}
& & 
& \texttt{com.kuaishou.nebula}	& Video	& -- & 390M & 540M & 23.2M && 15.7B & -- & -- & -- \\ \cline{4-14}
& & 
& \texttt{com.kwai.videoeditor}	& Video	& -- & 24.7M & 84.4M & 1.98M && 1.1B & -- & -- & -- \\ \cline{4-14}
& & 
& \texttt{com.kwai.livepartner}	& Video	& -- & 12.4M & 7.4M & 670K && -- & -- & -- & -- \\ \cline{4-14}
& & 
& \texttt{com.kwai.thanos}	& Video	& -- & 8.1M & 6M & 421K && 25M & -- & -- & -- \\
\hline 
3 & \texttt{a7a4ec5bc5} & VIPSHOP 
& \texttt{com.achievo.vipshop}	& Shopping	& -- & 690M & 45.8M & 6.7M && 14.4B & -- & -- & -- \\
\hline 
4 & \texttt{1315d03f03} & Ctrip
& \texttt{ctrip.android.view}	& Travel	& 1M+ & 380M & 32.1M & 92.8M && 10.6B & 1.4B & 3.0B & 210M \\
\hline 
5 & \texttt{b031fe98a4} & Xiaomi
& \texttt{com.xiaomi.smarthome}	& Tools	& 50M+ & 140M & -- & 1M && 1B & 7.1B & -- & --  \\ \cline{4-14}
& & 
& \texttt{com.xiaomi.wearable}	& Tools	& 10M+ & 542K & -- & -- && 9M & 400M & -- & -- \\ \cline{4-14}
& & 
& \texttt{com.xiaomi.router}	& Tools	& 5M+ & -- & -- & -- && -- & 100M & -- & -- \\ \cline{4-14}
& & 
& \texttt{com.xiaomi.mico}	& Tools	& -- & 7.7M & -- & 361K && 200M & 200M & -- & -- \\ \cline{4-14}
& & 
& \texttt{com.wali.live}	& Social	& -- & 446M & -- & -- && -- & -- & -- & -- \\ \cline{4-14}
& & 
& \texttt{com.xiaomi.superhexa}	& Tools	& -- & 1.3K & -- & -- && -- & -- & -- & -- \\
\hline 
6 & \texttt{14ceb70dfb} & Go Live 
& \texttt{com.gau.go.launcherex}	& Utility	& 100M+ & & && & & & & \\
\hline 
7 & \texttt{fdf6b1e025} & EastMoney
& \texttt{com.eastmoney.android.berlin}	& Finance	& -- & 24.6M & & 1.4M && 5M & -- & -- & -- \\ \cline{4-14}
& & 
& \texttt{com.eastmoney.android.newyork}	& Finance	& -- & 80K & 240K & 11K && 3M & -- & -- & -- \\
\hline 
8 & \texttt{509e03e376} & Baidu 
& \texttt{com.baidu.carlife}	& Tool	& -- & 3.7M & 1.5M & 524K && -- & -- & -- & -- \\
\hline 
9 & \texttt{d1e73a6359} & OneMicroWorld
& \texttt{com.yunio.smallworld}	& Social	& -- & 272K & -- & 5.6K && -- & -- & -- & -- \\
\hline
\end{tabular}
}
\begin{flushleft}
	\begin{footnotesize}
		\textsuperscript{$\dagger$} The data is retrieved from the public meta-data from the selected app stores as of January 2025. ``--'' indicates the app is not available on that store. \\
            \textsuperscript{$\ddagger$} The three alternative stores are (A1) Tencent App Store, (A2) Baidu App Store, and (A3) Wandoujia. The four OEM vendor app stores are owned by (O1) Huawei, (O2) Xiaomi, (O3) Oppo, and (O4) Vivo, respectively.\\
	\end{footnotesize}
\end{flushleft}
\end{table*}

%% file: tables/table_of_preinstalled_apps_2.tex
\begin{table}[t]
\centering
\def\arraystretch{0.98}
\setlength{\tabcolsep}{3pt}
\caption{\label{tab:table-compromised-pre-apps} List of pre-installed third-party apps involved in leakage of signing key with its fingerprint as \texttt{a6ef817bfd}}
\resizebox{1\linewidth}{!}{%
\begin{tabular}{HHllcHcc}
\hline

& \multirow{2}{*}{\textbf{\makecell[lb]{Certificate\vspace{-0.15em}\\Fingerprint}}} &  &  & \multirow{2}{*}{\textbf{\makecell[cb]{Involved\vspace{-0.15em}\\Vendor}}} & \multicolumn{3}{l}{\textbf{\makecell[lt]{Privilege Tier\textsuperscript{$\dagger$}}}} \\
\cline{6-8}
\textbf{\#} & & \textbf{App} & \textbf{Category} & & \textbf{Normal} & \textbf{System} & \textbf{Privileged} \\ \hline
1 & \texttt{a6ef817bfd} & \texttt{com.baidu.input} & \makecell[lt]{Input\\Method} & \vendorXiaomi &  \OK &   &   \\ \cline{5-8}
2 &   &   &   & \vendorVivo &   &   & \OK  \\ \cline{5-8}
3 &   &   &   & \vendorOppo &   & \OK  &   \\ \cline{3-8}
4 &   & \texttt{com.baidu.map.location} & Location & \vendorSamsung &   &   & \OK \\ \cline{5-8}
5 &   &   &   & \vendorVivo &   &   & \OK \\ \cline{3-8}
6 &   & \texttt{com.baidu.location.fused} & Location & \vendorSamsung &   &   & \OK  \\ \cline{5-8}
7 &   &   &   & \vendorHonor &   &  & \OK \\ \cline{3-8}
8 &   & \texttt{com.baidu.swan} & Utility & \vendorHonor &   & \OK &   \\
\hline
\end{tabular}%
}
\begin{flushleft}
\begin{footnotesize}
\textsuperscript{$\dagger$} System apps are pre-installed in the ``\texttt{system/app/}'' and ``\texttt{product/app/}'' directories. Privileged apps, on the other hand, are pre-installed in the ``\texttt{system/priv-app/}'' directory. Any app that is neither a system app nor a privileged app is typically installed in the ``\texttt{data/app/}'' directory.\\
\end{footnotesize}
\end{flushleft}
\end{table}

%% file: tables/table_of_aosp_key_usage_2.tex
\begin{table}[t]
\centering
\def\arraystretch{0.98}
\setlength{\tabcolsep}{3pt}
\caption{\label{tab:table-aosp-key-usage} List of public AOSP keys that have been signed in real-world apps}
\footnotesize
\resizebox{1\linewidth}{!}{%
\begin{tabular}{HlllH}
\hline

& \multirow{2}{*}{\textbf{\makecell[lb]{Certificate\vspace{-0.25em}\\Fingerprint}}} &  & \multirow{2}{*}{\textbf{\makecell[lb]{Involved\vspace{-0.25em}\\OEM Vendors}}} &  \\
\textbf{\#} & & \textbf{App} & & \textbf{Note} \\ \hline
1 & \texttt{097b2bd6d4} & \texttt{com.android.nearby.halfsheet} &  \vendorMeizu &    \\ \hline 
2 & \texttt{366f75e331} & \texttt{com.android.safetycenter.resources} & \vendorMeizu, \vendorOppo, \vendorVivo & \\ \hline
3 & \texttt{43ddcf4e8a} & \texttt{com.android.uwb.resources} & \makecell[lt]{Pixel, \vendorMeizu, \vendorOppo,\\\vendorVivo} & \\ \hline
4 & \texttt{6ef8a09a1c} & \texttt{com.android.wifi.resources} &  \makecell[lt]{\vendorHonor, \vendorHuawei, \vendorMeizu,\\ \vendorOppo, \vendorVivo} &    \\ \hline 
5 & \texttt{936ef21771} & \texttt{com.android.hotspot2.osulogin} &  \makecell[lt]{\vendorHonor, \vendorHuawei, \vendorMeizu,\\ \vendorOppo, \vendorVivo} &    \\ \hline 
6 & \texttt{a6ccc500ff} & \texttt{com.android.bluetooth} &  \vendorVivo &    \\ \hline 
7 & \texttt{abf21f9e2a} & \texttt{com.android.sdksandbox} &  \vendorMeizu, \vendorVivo & Undecrypted   \\ \hline 
8 & \texttt{c97c5176ed} & \texttt{com.android.connectivity.resources} &  \makecell[lt]{\vendorHonor, \vendorHuawei, \vendorMeizu,\\\vendorOppo, \vendorVivo} &    \\ \hline 
9 & \texttt{ccc51124be} & \texttt{com.android.adservices.api} &  \vendorMeizu, \vendorOppo, \vendorVivo &    \\ \hline 
10& \texttt{dc79bd2900} & \texttt{com.android.wifi.dialog} &  \vendorMeizu, \vendorOppo, \vendorVivo &    \\ \hline
11& \texttt{e1dbadce60} & \texttt{com.android.captiveportallogin} &  \vendorMeizu &    \\ \cline{3-5}
 & \texttt{} & \makecell[lt]{\texttt{com.android.cellbroadcastreceiver.}\,\\\texttt{module}} &  \vendorMeizu &    \\ \cline{3-5}
 & \texttt{} & \texttt{com.android.cellbroadcastservice} &  \vendorMeizu &    \\  \cline{3-5}
 & \texttt{} & \texttt{com.android.networkstack} &  \vendorMeizu &    \\ \cline{3-5} 
 & \texttt{} & \texttt{com.android.networkstack.tethering} &  \vendorMeizu &    \\ \hline
12& \texttt{b031fe98a4} & \texttt{com.xiaomi.wearable} &  \vendorXiaomi &    \\ \cline{3-5}
& \texttt{} & \texttt{com.xiaomi.smarthome} & \vendorXiaomi &    \\ \cline{3-5} 
& \texttt{} & \texttt{com.xiaomi.superhexa} & \vendorXiaomi &    \\ \cline{3-5} 
& \texttt{} & \texttt{com.xiaomi.mico} &  \vendorXiaomi &    \\ \cline{3-5} 
& \texttt{} & \texttt{com.xiaomi.router} & \vendorXiaomi  &    \\ \cline{3-5} 
\hline
\end{tabular}%
}
\end{table}

%% file: sections/exploit.tex
\section{Exploitation (RQ4)} \label{sec:exploit}

Leaked app-signing keys can be exploited by malicious parties.
They can forge apps on behalf of the developer if they manage to recover both the passwords for the keystore and the corresponding private key.
Those forged apps can evade all signature verification mechanisms in the Android ecosystem, and more seriously, can launch attacks if the victim device installs them as an update of the previously installed benign apps. 

\subsection{Attack Strategies} 
Our proposed exploitation is conceptually inspired by the approach of Janus vulnerability (CVE-2017-13156)~\cite{sanatsu2021exploit}. 
Unlike that Janus exploit, which requires extensive analysis of the APK structure, our attack does not rely on parsing the signature content but only needs to repack the app by signing with the leaked keys one more time. 

\begin{figure*}%
	\begin{minipage}[b]{0.4\linewidth}
		\centering
		\includegraphics[width=1\linewidth]{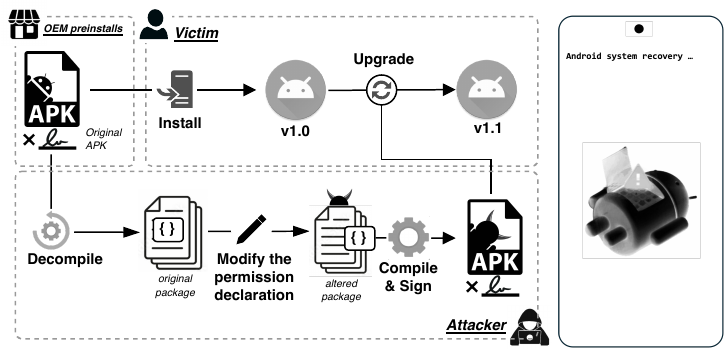}\\
		\vspace{0.6cm}
		\subcaption{}
	\end{minipage}%
	\hspace{1em}
	\begin{minipage}[b]{0.45\linewidth}
		\centering
		\includegraphics[width=1\linewidth]{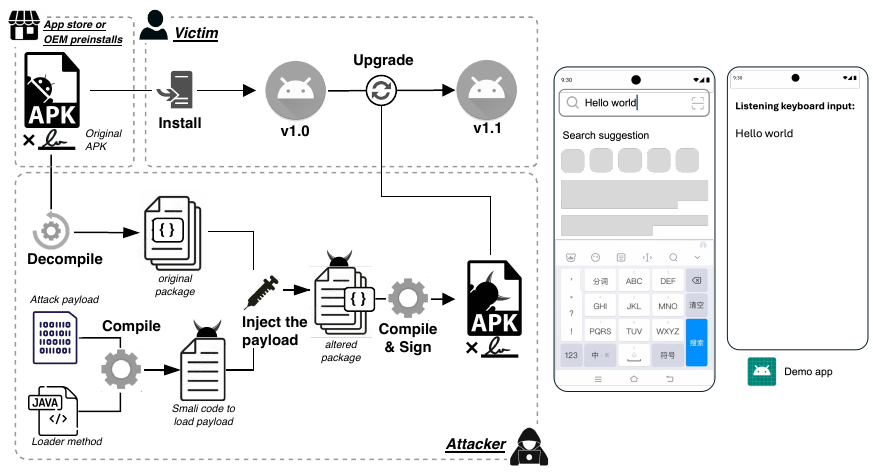}\\
		\vspace{-0.1cm}
		\subcaption{}
	\end{minipage}%
	\vspace{-0.2cm}
	\caption{The process of (a) creating permission conflict and (b) injecting malicious DEX.}
	\label{fig:exploit}
\end{figure*}

To realize malicious manipulation in the repacked apps, we propose two strategies for modifying the original app packages:

\begin{enumerate}[wide]
\item \textbf{Creating permission conflict}.
Considering many victim apps have been found privileged in OEM devices, altering the permission declaration in these apps would result in permanent denial of service of the victim device. We depict this process in Figure~\ref{fig:exploit}~(a).
Specifically, the attacker can reverse engineer a privileged app signed with the compromised certificate, and in the forged version intentionally requests more permissions than what the benign app has been explicitly granted in the allowlist\footnote{Allowlists are implemented in \texttt{.XML} files in the \texttt{frameworks/base/etc/permissions} directory~\cite{stackoverflow2025root}.}. By doing this, the Android OS detects an inconsistency in handling privileged-level permissions on the next boot, and as a result, will keep crashing permanently~\cite{google2025allowlisting}. 
\item \textbf{Injecting malicious DEX}.
The Android OS allows developers to insert and execute pre-compiled DEX files into the apps~\cite{google2025inmemorydex}. This feature lays the technical foundation to support headless plugins for the millions of Android apps~\cite{bisset2023exploring,stackoverflow2010inject} and also enables the creation of the Janus vulnerability exploitation~\cite{sanatsu2021exploit}. 
We can also inject maliciously implemented DEX files into the original apps to abuse them.
Specifically, we can first implement an attack script in Java, which usually performs network sniffing, user input monitoring, or remote execution.
We then compile the attack script to DEX format to obtain an attack payload, and write a ``loader'' method to load the DEX payload during runtime. 
An example of the loader method is provided in Listing~\ref{lst:loader}.
After that, we compile the loader method together with the payload into Smali code\footnote{Smali is an assembly language used by Android, which can be used to modify and reverse engineer Android apps, and allows developers to make changes to the bytecode of an app~\cite{kumar2023smali}.} and insert it into the entry point of the begin app, i.e., when the app initializes. 
Lastly, a spoofed version of the victim app can be made by re-signing the modified package using the leaked keys.
Thus, our prepared attack payload can be executed each time the spoofed app is launched on the victim device.
We depict this process in Figure~\ref{fig:exploit}~(b) and implement a proof-of-concept based on this strategy, which will be detailed in the remainder of this section.
\end{enumerate}

\subsection{A Proof-of-Concept}

We begin by simulating a typical hacker's approach in building a malicious app, with an assumption that the original app can be reverse-engineered and subsequently re-built with its source code having been tampered with.
In this study, we roughly analyze the functionality of the \noofVulnerableApps compromised apps and use a popular reverse engineering tool named \texttt{ApkTool}~\cite{tumbleson2025apktool} to assess their exploitability, i.e., whether we can rebuild the decompiled app package. 
For demonstration purposes, we aim to build a forged app with the least code modification, and to the maximum extent, avoid the forged app from triggering any potential security mechanisms like online verification and file integrity checks. 
For those reasons, we exclude the apps that excessively request an Internet connection to function, and as a result, we selected the Baidu input method as the victim app.

We consider the victim app is granted with access to the user's keyboard input and accordingly craft an attack script to read and save users' keyboard input into a file located in the public storage (i.e., the \texttt{/sdcard} directory).
We then compile the script and receive the attack payload as a DEX file.
Next, we write a loader method to load the attack payload at runtime and invoke the attack methods. We compile the loader to Smali code, and inject it into the place where the victim app is initialized after each time the app is launched. 
Lastly, we use the compromised key to sign the modified package with a higher version\footnote{We build the forged app with version \texttt{v11.6.12.17} to upgrade the benign one \texttt{v.11.6.12.5}}, and then we obtain the forged app.

\input{sections/algorithm_exploit}

To facilitate demonstration, we also write an auxiliary app to actively read the saved keyboard input from the storage. Thus, after the user installs the forged app (i.e., through app update), any keyboard input using the forged Baidu input method will be stealthily recorded and reflected in the auxiliary app in real time, as illustrated in Figure~\ref{fig:exploit}~(b). 
The key implementations of this proof-of-concept, together with a demonstration video, can be found in our anonymous repository (see Section~\ref{sec:data_availability}).

We note that this proof-of-concept only reflects the exploitability of our findings at the moment this research was conducted. 
The spoofing attack can be theoretically applied to any compromised app, given that the attacker has sufficient toolkit and reconnaissance skills. However, the developer of the victim app may strengthen its security mechanism to proactively prevent possible attacks in future releases.   

%% file: sections/algorithm_exploit.tex

\begin{figure}[t]
\begin{lstlisting}[caption={Code for DEX injection used in our proof-of-concept exploit},captionpos=b,language=Java,label={lst:loader}]
public static void loadMemoryDex(String payload, String className, String methodName, String params) {
	try {
		// Initiate an InMemoryDexClassLoader object using the attack payload
		Object dexClassLoaderObj = Class.forName("dalvik.system.InMemoryDexClassLoader")
			.getDeclaredConstructor(ByteBuffer.class, ClassLoader.class)
			.newInstance(ByteBuffer.wrap(Base64.getDecoder().decode(payload)), Object.class.getClassLoader());
		// Initiate the findClass method
		Method findClassMethod = Class.forName("dalvik.system.BaseDexClassLoader")
			.getDeclaredMethod("findClass", String.class);
		findClassMethod.setAccessible(true);
		// Load the attack method defined in the payload
		Class<?> dexTaskClass = (Class<?>) findClassMethod.invoke(dexClassLoaderObj, className);
		// Execute the attack by direct invocation
		dexTaskClass.getDeclaredMethod(methodName, String.class, String.class).invoke(null, params);
	} catch (Exception e) {
		e.printStackTrace();
	}
}
\end{lstlisting}
\end{figure}

%% file: sections/broader_impact.tex
\section{Broader Impact beyond the Mobile Platforms}
\label{sec:broader-impact}

\begin{figure}[t]
	\centering
	\vspace{-0pt}
	\includegraphics[trim=0cm 0cm 0cm 0cm, clip=true,width=1\linewidth]{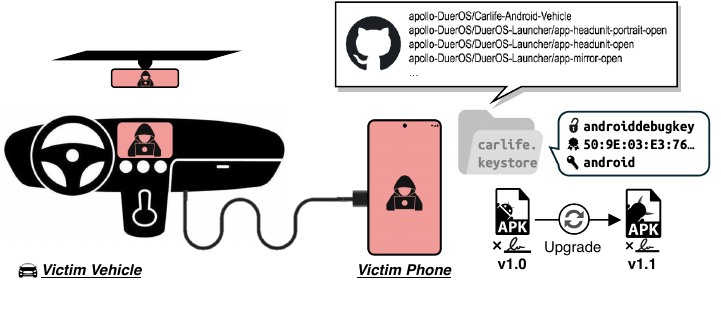}
	\vspace{-0.5cm}
	\caption{An illustrated attack to hijack the automotive head unit in the vehicle through the victim's smartphone after the private key being compromised}
	\label{fig:carlife-issue}
	\vspace{-0.1cm}
\end{figure}

A compromised keystore named ``\texttt{carlife.keystore}'' in a repository ``\texttt{Carlife-Android-Vehicle}'' drew our attention when we sorted our findings. Although we did not find an app matching the signing key, it still raised serious concern because its filename links to a popular vehicle platform owned by Baidu. 
The same keystore can also be found in our findings with filename ``\texttt{debug.keystore}'' in both the same project and another repository named ``\texttt{DuerOS-Launcher}''.

We then performed a dedicated investigation of the found keystore to assess the potential risk and impact. Specifically, we searched for candidate apps designed either for the Android mobile platform or for the vehicle platform. We found a public free app named ``CarLife+'' that can only be downloaded through the Baidu CarLife official website. By extracting the APK's signature, we managed to match the SHA-256 digest of the certificate with the private key included in the compromised keystore. We also learned that the Baidu Carlife is one of the popular automotive head unit platform with the largest user base in China, and the DuerOS is the vehicle-to-everything (V2X) OS that works with Baidu CarLife.  
According to these findings, we conclude that the signing key for CarLife has been leaked and the impact of the signature leakage incident has extended to smart vehicle platforms.
An attacker can take advantage of this compromised keystore to launch a spoofing attack by replacing the original CarLife app on the victim's phone, and therefore the fake CarLife app can collude with other malicious apps on the device to monitor users' privacy and driving behavior. Even without a colluding app, the fake CarLife app can still interrupt the normal operation of the automotive head unit in the vehicle and conduct a DoS attack, such as injecting pop-up Ads or black screen, as conceptually illustrated in Figure~\ref{fig:carlife-issue}. Similar attack approaches can pose even more severe safety risks in future generations of smart vehicles when autonomous driving becomes more prevalent.   

We note that the potential impact of the signature leakage issues in the Android ecosystem might be more serious than on other platform because the Android OS has become widely installed on not only mobile devices but also vehicles, IoT, home appliance, and diverse smart home hardware. 
The affected population may be unexpectedly large, and the mitigation could be an extremely costly and lengthy process.
Take CarLife as an example, it has been installed on more than 1,100 vehicle models of 99 manufacturers in the Chinese market, including the global best-selling ones like Toyota, Ford, Hyundai, Audi, BMW, Volkswagen, and Mercedes, dominating the automotive head unit market share in China~\cite{baidu2025carlife}.

%% file: sections/disclosure.tex
\section{Ethical Disclosure}
\label{sec:ethic}

We responsibly disclosed our confirmed findings to the relevant parties and withheld case-specific details for at least 90 days prior to publication. For affected third-party apps, we manually identified developer contact channels and reported the observed signing-key exposure and associated risks. Remediation progress remained uneven at the time of writing. While some developers acknowledged the reports, only one third-party developer had completed key rotation among the cases listed in Table~\ref{tab:table-compromised-apps}. We also observed that the repositories involved in these confirmed cases were no longer publicly accessible, which reduces immediate re-exposure risk, although it does not necessarily imply full remediation of the affected signing identities.

%% file: sections/discussion2.tex
\section{Recommendations} \label{sec:recommendation}

Our findings point to recurring weaknesses in the handling of Android signing credentials across both third-party development and OEM deployment settings. Based on these observations, we offer several practical recommendations to reduce the risk of signing-key exposure and limit its downstream consequences.

\subsubsection*{\textbf{Strengthen Credential Management}}
Developers should treat app-signing credentials as long-lived security-critical assets rather than ordinary build artifacts. Keystore passwords should never be stored in plaintext within project files or repository-accessible paths, and weak passwords should be avoided.
For Android developers, we further recommend using distinct passwords for the keystore and the signing key to reduce the risk of compromise.
We also advocate that developers store signing credentials in environments protected by multi-factor access control to minimize the risk of unexpected exposure. 

\subsubsection*{\textbf{Delegate App Signing to Trusted Services}}
To reduce the risk of key exposure, developers should consider delegating signing operations to trusted managed services, such as Google's Play App Signing infrastructure~\cite{google2023sign}. Such services can isolate private keys from local development workflows, and support recovery when developer-side environments are compromised. This is particularly beneficial for teams that do not have mature internal processes for secret management and release hardening.

\subsubsection*{\textbf{Enforce Stricter App Signing Discipline}}
Device manufacturers (i.e., OEMs) should never sign system apps using public platform keys provided by Google or any other shared certificate authority. 
OEMs should sign OS images and critical system components in an isolated environment and refrain from pre-installing third-party apps within the OS image. 
More importantly, they should never sign third-party apps with their own system keys, as this grants excessive privileges and creates additional opportunities for abuse in the event of a compromise. 
Such practices can turn an app with known vulnerabilities into a vector for privilege escalation attacks.

\section{Threats to Validity} \label{sec:limitation}

Our methodology may have a few internal constraints due to the design choices and scope boundaries. Our work is also subject to several external limitations outside our control.

\subsection{Internal Limitations}

\subsubsection*{\textbf{Repository and File-Format Scope}}
Our repository-side analysis is centered on Android projects and focuses on the default Java keystore formats commonly used in Android release workflows, namely \texttt{.jks} and \texttt{.keystore}. This design supports a consistent and reproducible extraction pipeline, but it may miss signing materials stored in non-Android repositories or in other formats such as \texttt{.pem} and \texttt{.p12}. As a result, our measurements may understate the broader landscape of signing-key exposure.


\subsubsection*{\textbf{Password Leakage Criteria}}
We conservatively classify a key as compromised only when the corresponding plaintext password appears directly in the project configuration or can be resolved through an accessible file path referenced by the repository. Our study does not consider more complex password-management practices, encrypted storage, or organization-specific signing workflows. In addition, although some exposed passwords appear weak, we do not attempt password cracking for ethical and legal reasons. These choices likely make our analysis underestimate the scale of weak credentials in the real world.

\subsubsection*{\textbf{Repository Coverage}}
Our analysis focuses on public GitHub repositories because of their scale and relevance to modern software development~\cite{lambropoulos2022github}. However, Android projects may also be hosted on other platforms, such as GitLab, Bitbucket, and Gitee. Leakage patterns on those platforms may differ, which limits the generalizability of our measurements to the broader repository ecosystem.

\subsection{External Limitations}

\subsubsection*{\textbf{App Source Selection}}
Our third-party app collection relies on public app-store rankings and retrieves APK files through APKPure, primarily to support a lightweight and reproducible pipeline for obtaining the latest app versions. We acknowledge that academic app archives such as AndroZoo~\cite{allix2016androzoo} provide a valuable large-scale resource by mirroring Android apps from multiple sources, and could also serve as an alternative source of APKs for similar studies. 

After verifying the affected third-party apps identified in this work, we found that 24 out of the 26 compromised third-party apps can also be located in AndroZoo. This suggests that our findings are unlikely to be driven solely by the choice of APKPure as the download source. Nonetheless, the exact availability and version alignment of APKs may vary across providers, which may affect the completeness and reproducibility of app collection in future replications.

\subsubsection*{\textbf{Search and Retrieval Constraints}}
Systematic large-scale collection from GitHub is affected by evolving search restrictions and limitations on external tool access. Although we adopted practical workarounds, these constraints may reduce the completeness and reproducibility of data collection.

\subsubsection*{\textbf{Temporal Limitation}}
Our dataset is derived from snapshots collected during the study period rather than from a complete historical archive of repository states. Developers may subsequently add, remove, or rotate signing keys. Therefore, the reported landscape should be interpreted as a conservative snapshot and may not capture all historical leakages or emerging leakage patterns.

\subsubsection*{\textbf{Exploitation Complexity}}
While we demonstrate a proof-of-concept attack using a compromised key, the practical impact of such attacks depends on various factors, including the app's user base, the attacker's reverse engineering capabilities, and the feasibility of crafting an attack payload (e.g., triggering privileged operations and/or exfiltrating sensitive data). The lack of a universal push-button-like automated pipeline limits the real-world applicability of exploitation at scale.

\subsubsection*{\textbf{Ethical Boundaries}}
We intentionally refrain from password cracking to ensure our methodology aligns with ethical norms. However, this may cause us to miss additional exploitable cases.

%% file: sections/related_work.tex
\section{Related Work}

\subsubsection*{\textbf{Android App Signing and Developers' Practice}}
The security of Android's app signing mechanism has been the subject of multiple studies, although there is still no comprehensive large-scale measurement of the security landscape through the lens of developer key leakage.
Early research by Zhou and Jiang~\cite{zhou2012dissecting} explored malicious Android applications, highlighting that attackers could exploit poorly managed signing keys to deploy malware under legitimate identities. Similarly, studies by Barrera \textit{et al.}~\cite{barrera2010methodology} and Felt \textit{et al}.~\cite{felt2011android} evaluated permission models and developer practices, noting that security vulnerabilities often stem from developers' insufficient security knowledge and poor credential management.

Relevant work, such as Egele \textit{et al}.~\cite{egele2013empirical}, assessed the Android ecosystem's robustness by analyzing apps' cryptographic practices, concluding that many apps suffered from severe cryptographic misuse or poor key management practices. 
Wang \textit{et al.}~\cite{wang2019characterizing} comprehensively investigated the security mechanism of early versions of Android app signing schemes and proposed a taxonomy covering different types of app signing issues.
The authors of~\cite{wang2019characterizing} also indicate that some amateur Android developers use publicly known keys to sign their apps. Once such keys are leaked, attackers can craft malicious apps bearing authentic signatures.
More recently, Wei \textit{et al.}~\cite{wei2025far} performed a systematic assessment of secret leakage within the app packages and identified over 3,700 exploitable app secrets, underscoring the worrisome security practices in the Android developer community.
Shi \textit{et al.}~\cite{shi2025skeleton} studied credential leakages of mini-apps on their hosting super-app platforms, uncovering over 84,000 credential leaks spanning over 54,000 mini-apps. 

To the best of our knowledge, this is the first study that examines the Android signing key mishandling at scale and systematically investigates the impact in the broader ecosystem.

\subsubsection*{\textbf{Mining Software Repositories for Credential Leakage and Mobile Security}}

There is a growing body of work that applies mining software repositories techniques to uncover credential and key leakage issues.
Meli \textit{et al.}~\cite{meli2019bad} conducted a large-scale longitudinal analysis of billions of files across GitHub and searched for private RSA key leakage from them, showing that hard-coded passwords and API keys were among the most prevalent causes of exposure.
Sinha \textit{et al.}~\cite{sinha2015detecting} proposed a hybrid detection approach combining keyword search, pattern matching, heuristic filtering, and code slicing to pinpoint API keys from the ocean of files hosted on GitHub.
Shi \textit{et al.}~\cite{shi2021empirical} studied Android mobile payment credential leakage by mining public Git repositories and demonstrated how attackers can exploit them to forge package signatures and hijack payment flows in third‑party cashier systems.
Feng \textit{et al.}~\cite{feng2022automated} proposed PassFinder, a deep learning based system to explore password leakage in GitHub repositories. It leverages deep neural networks to unveil the intrinsic characteristics of textual passwords and comprehend the code semantics that use these passwords for authentication purposes.

Existing research on repository mining has showcased mature techniques for detecting credential leakage by inspecting the code base. 
Our research builds on the idea of mining software repositories and complements existing literature with a focus on app signing as part of release engineering in Android app development lifecycle.

%% file: sections/conclusion.tex
\section{Conclusion} \label{sec:conclusion}

In this work, we investigate the security landscape of app signing in the Android ecosystem with a particular focus on developer-side key management. We conduct a large-scale measurement covering \noUniqueKs keys collected from the open repositories and \noAppsSig developer signatures extracted from over 4000 apps.
By recovering passwords exposed by repository owners, our analysis finds 26 compromised certificates threatening the security of 278 apps and affecting over 10 billion users. 
We also implement a proof-of-concept attack on a real device to demonstrate the exploitability. 
Our further analysis reveals that the impact of signing key leakage already extends beyond mobile devices and threatens the security of vehicles.
Our findings should raise an alert to both developers and OEM manufacturers to comply with rigorous security practices throughout the entire development life cycle and urge the industry to adopt a more secure key management solution. 


%% file: sections/data_availability.tex
\section{Data Availability Statement} \label{sec:data_availability}

Our artifacts are made available for anonymous review at \repoDoi (or at \repoUrl). 
We have also uploaded a demonstration video of the proof-of-concept exploit to the repository.